# Magnetoelectric Effects in Ferromagnetic Metal-Piezoelectric Oxide Layered Structures

U. Laletsin[1], N. Paddubnaya[1], G. Srinivasan[2], M.I. Bichurin[3]

[1]*Institute of Technical Acoustics, National Academy of Sciences of Belarus, 210717 Vitebsk, Belarus*
[2]*Physics Department, Oakland University, Rochester, Michigan, USA*
[3] *Department of Engineering Physics, Novgorod State University, Novgorod, Russia*

**ABSTRACT**

Frequency dependence of magnetoelectric (ME) coupling is investigated in trilayers of ferromagnetic alloy and piezoelectric lead zirconate titanate (PZT). The ferromagnetic phases studied include permendur, a soft magnet with high magnetostriction, iron, nickel, and cobalt. Low frequency data on ME voltage coefficient versus bias magnetic field indicate strong coupling only for trilayers with permendure or Ni. Measurements of frequency dependence of ME voltage reveal a giant ME coupling at electromechanical resonance. The ME interactions for transverse fields is an order of magnitude stronger than for longitudinal fields. The maximum voltage coefficient of 90 V/cm Oe at resonance is measured for samples with nickel or permendure and is three orders of magnitude higher than low-frequency values.

## 1. INTRODUCTION

Harshe and co-workers proposed a new direction in the development of ME composites, i.e., magnetostrictive-piezoelectric layerd structures [1]. The primary advantages of such structures are low leakage currents and a high degree of polarization. The use of ferromagnetic metals helps to accomplish a very high degree of polarization compared to oxide ferromagnets [2]. This work is devoted to investigations of ME effects in a trilayer metal-piezoelectric-metal structure. We used lead zirconate titanate (PZT) for the piezoelectric phase. The thickness of PZT layers varied from 0.3 to 0.8 mm. Permendure, a Ni-Fe alloy, Fe, Ni or Co with a thickness of 0.18 mm was used as the ferromagnetic phase. Trilayers were made by bonding the PZT and the metal disks.

The ME voltage coefficient was determined by measuring the electric field generated across the piezoelectric layer when an a.c. magnetic field and a d.c. magnetic field were applied to sample. There were two variants of the experiment. In the first configuration, the electrical polarization was perpendicular to magnetic fields (transverse ME effect). In the second, electrical polarization was parallel to magnetic fields (longitudinal ME effect). The measurements were carried out at low frequencies (1.0 kHz) and at electromechanical resonance frequency (~330 kHz). The resonance frequency was determined by sample size and composite parameters.

The d.c magnetic field dependence of ME voltage coefficient (MEVC) was measured for transverse and longitudinal fields. We measured an increase in MEVC with increasing H, followed by a drop at high fields. The magnitude of MEVC was found to depend on the thickness and nature of the bonding medium. The maximum MEVC was measured for samples with PZT thickness of 0.4 mm and 0.36 mm. Longitudinal ME effect was an order of magnitude smaller than transverse ME effect. The maximum ME voltage coefficient obtained at low frequency varied from 0.2 to 0.65 V/cm Oe, depending on the nature of ferromagnetic phase. Samples with Co yielded the lowest MEVC and the highest are for trilayers with permendur. The maximum ME voltage coefficient obtained at resonance frequency was equal to 90 V/cm Oe. These results are analyzed in terms of recent theoretical models for ME effects in bilayers [3].

## 2. Experiment

Trilayer structures of ferromagnetic permendur, iron, nickel, or cobalt and piezoelectric PZT were synthesized. Permendur (P) is a soft magnetic alloy consisting of 49% iron, 49% cobalt and 2% vanadium. It is an ideal material for studies on ME composites due to desirable low resistivity and high magnetization (23.4 kG), Curie temperature (1213 K), permeability and magnetostriction (70 ppm) [4]. PZT was chosen due to high ferroelectric Curie temperature and piezoelectric coupling constant. PZT, 9 mm in diameter, was first poled and then bonded to metals with 0.01-0.03 mm thick layer of an epoxy. Trilayers were made with the central PZT layer bonded to outer metal layers. For P-PZT-P, the thickness of metal layer was 0.18 mm for all the samples studied. PZT thickness varied from 0.2 mm to 0.8 mm. For samples with transition metals, the thickness of metals and PZT was 0.4 mm.

For ME characterization, we measured the electric field produced by an alternating magnetic field applied to the biased composite. The samples were placed in a shielded 3-terminal holder and placed between the pole pieces of an electromagnet that was used to apply the bias magnetic field H. The required ac magnetic field of $\delta H$=0.01-1 Oe at 10 Hz - 1 MHz parallel to H was generated with a pair of Helmholtz coils. The induced electric field $\delta E$ perpendicular to the sample plane was estimated from the voltage $\delta V$ measured with an oscilloscope. The ME voltage coefficient is estimated from $\alpha_E = \delta E/\delta H = \delta V/t\, \delta H$ where t is the thickness of PZT. The measurements were done for two different field orientations. With the sample plane represented by (1,2), the transverse coefficient $\alpha_{E,31}$ was measured for the magnetic fields H and $\delta H$ along direction-1 (parallel to the sample plane) and perpendicular to $\delta E$ (direction-3). The longitudinal coefficient $\alpha_{E,33}$ is measured for all the fields perpendicular to the sample plane. Magnetoelectric characterization was carried out at room temperature as a function of frequency of the ac magnetic field and bias magnetic field H.

## 3. Results and Discussion

Figure 1 shows representative data on the H dependence of the longitudinal and transverse ME voltage coefficients. The measurements were done at 1 kHz on a P-PZT-P trilayer sample with a PZT thickness of 0.36 mm. For the longitudinal fields, as H is increased from zero, one observes increase in $\alpha_{E,33}$ to a maximum at $H_m$ = 600 Oe. With further increase in H, $\alpha_{E,33}$ decreases rapidly to a minimum. When H is reversed, $\alpha_{E,33}$ becomes negative (a 180 deg. phase difference between $\delta H$ and $\delta E$), but the field dependence remains the same as for positive H. For transverse fields, Fig.1 shows a similar variation in the ME voltage coefficient with H as for the longitudinal case but with the following departures. (i) The maximum in $\alpha_{E,31}$ occurs at a much smaller $H_m$ of 150 Oe compared to the longitudinal case. (ii) The peak value of $\alpha_{E,31}$ is an order of magnitude higher than for $\alpha_{E,33}$.

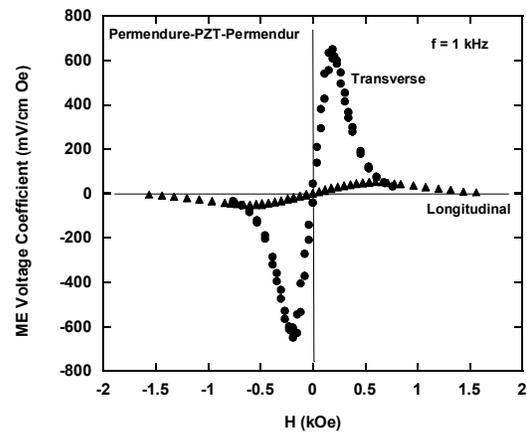

*Fig.1: Variations in longitudinal and transverse magnetoelectric (ME) voltage coefficients with the bias magnetic field H for a trilayer composite of Permendur-PZT-Permendur.*

Similar measurements were made on trilayers of Fe-PZT-Fe, Co-PZT-Co and Ni-PZT-Ni and the data for transverse MEVC are shown in Fig.2. From the data, one infers the presence of strong ME coupling. The coupling strength, however, varies with the nature of ferromagnetic metal, with the weakest coupling for Co containing samples and the strongest for trilayers with Ni. One also observes unique characteristics such as weak ME coupling for low fields and zero-crossing at high fields for Fe-PZT-Fe. Next we compare $\alpha_E$ values in Fig.1 and 2 with results for similar composites. Systems of interest in the past were bulk samples of ferrites with barium titanate or PZT. The $\alpha_E$ in Fig.1 are a lot higher than values reported for bulk composites of cobalt ferrite

(CFO) or nickel ferrite (NFO) with PZT or $BaTiO_3$ [1]. Layered composites studied so far include ferrite-PZT [2,3] lanthanum manganite-PZT [5] and terfenol-PZT [6]. For comparison, the highest value for $\alpha_{E,31}$ is 60 mV/cm Oe in bilayers of lanthanum manganites-PZT, 400 mV/cm Oe for NFO–PZT and 4680 mV/cm Oe for Terfenol-PZT. Thus the results in Fig.1 provide clear evidence for one of the highest ME coupling reported for any composites.

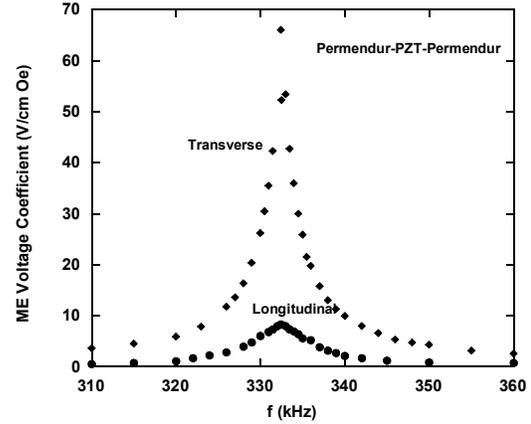

*Fig.3: Frequency dependence of transverse and longitudinal ME voltage coefficients for the P-PZT-P trilayer. The bias field H was set for maximum ME coupling (Fig.1). The resonance frequency corresponds to the electromechanical (EMR) for PZT in the composite.*

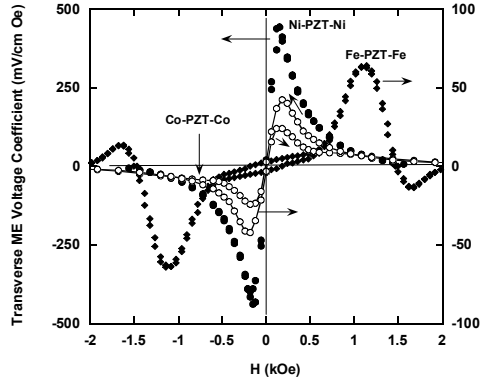

*Fig.2: Similar data as in Fig.1 for Fe-PZT-Fe, Co-PZT-Co and Ni-PZT-Ni.*

We also performed studies on the frequency dependence of the ME coupling. The bias field was set at $H_m$ and the voltage coefficients were measured as the frequency f of the ac field $\delta H$ was varied. Typical $\alpha_E$ vs f profiles for longitudinal and transverse fields are shown in Fig.3 for P-PZT-P. The results are for the trilayer with a PZT thickness of 0.36 mm. Consider first the data for the longitudinal field. Upon increasing f, $\alpha_{E,33}$ remains small and constant for frequencies up to 250 kHz. At higher f, we observe a rapid increase in $\alpha_{E,33}$ to a maximum of 8000 mV/cm Oe at 330 kHz. Finally, $\alpha_{E,33}$ levels off at 50 mV/cm Oe at high frequencies. The profile thus shows resonance with $f_r$ = 330 kHz and a width $\Delta f$ = 6 kHz, corresponding to a quality factor Q = 55. Figure 3 also shows a similar resonance in $\alpha_{E,31}$ for transverse fields. The resonance occurs at the same frequency as for the longitudinal fields, but with a much higher maximum $\alpha_E$ (=66000 mV/cm) Oe and a higher Q (=155) compared to the longitudinal fields. The transverse coupling is an order of magnitude stronger than the longitudinal voltage coefficient.

We carried out measurements of resonance ME effect in trilayers of Fe-PZT-Fe, Co-PZT-Co and Ni-PZT-Ni and the results for transverse MEVC are shown in Fig.4. A sharp resonance with a high Q is evident for all three samples and the resonance value of $\alpha_{E,31}$ follows the trend observed at low frequencies. A maximum value of 90 V/cm Oe at resonance is measured for both P-PZT-P and Ni-PZT-Ni.

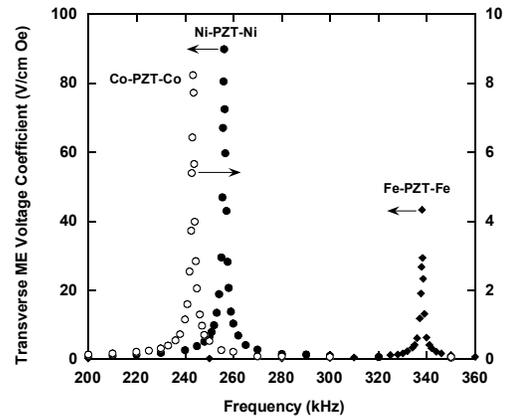

*Fig.4: Similar data as in Fig.3 for metal-PZT trilayers.*

Now we analyze the data in terms of theoretical models we developed recently for low frequency and resonance ME effects [7]. Our model predicts giant magnetoelectric interactions in ferromagnet-PZT bilayers at frequencies corresponding to electromechanical resonance. We considered a bilayer in the form of thin disk of radius *R*. The ac magnetic field induces harmonic waves in the radial or thickness modes. The model considers radial modes for transverse or longitudinal fields. An averaging procedure was employed to obtain the composite parameters and the ME voltage coefficient $\alpha_E$. The frequency dependence of $\alpha_E$ shows a resonance character at the electromechanical resonance for PZT in the bilayer. The resonance frequency depends on *R*, mechanical compliances, density and the coefficient of electromechanical coupling for radial mode. The peak value of $\alpha_E$ and the width of resonance are determined by the effective piezomagnetic and piezoelectric coefficients, compliances, permittivity and loss factor. Based the model, one expects a resonance in $\alpha_E$ versus frequency profile with a maximum $\alpha_E$ that is a factor of 40-1000 higher than low frequency values, depending on the nature of the magnetostrictive phase

In conclusion, we reported the observation of theoretically predicted giant ME interactions at electromechanical resonance in ferromagnetic alloy-PZT layered samples. The phenomenon could be utilized to accomplish very high field conversion efficiency in the product property composites.

The work at Oakland University was supported by a National Science Foundation grant (DMR-0302254).

### *References*

1. Harshe, J. P. Dougherty and R. E. Newnham, *Int. J. Appl. Electromag. Mater.* **4**, 145 (1993); M. Avellaneda and G. Harshe, *J. Intell. Mater. Sys. Struc.* **5**, 501 (1994).
2. Srinivasan, E. T. Rasmussen, J. Gallegos, R. Srinivasan, Yu. I. Bokhan, and V. M. Laletin, *Phys. Rev. B* **64**, 214408 (2001).
3. G. Srinivasan, E. T. Rasmussen, and R. Hayes, *Phys. Rev. B* **67**, 014418 (2003).
4. Ferromagnetism, R. Bozorth, IEEE Press (New York), 1993.
5. G. Srinivasan, E. T. Rasmussen, B. J. Levin, and R. Hayes, *Phys. Rev. B* **65**, 134402 (2002).
6. J. Ryu, A. V. Carazo, K. Uchino, and H. Kim, *Jpn. J. Appl. Phys.* **40**, 4948 (2001).
7. M. I. Bichurin, V. M. Petrov, and G. Srinivasan, *J. Appl. Phys.* **92**, 7681 (2002).